%% file: arxiv.tex
\pdfoutput=1
\documentclass[letterpaper,onecolumn,10pt]{article}
\usepackage[T1]{fontenc}
\usepackage[margin=1in]{geometry}
\usepackage{endnotes}
\usepackage{graphicx}
\usepackage{amsmath}
\usepackage{amssymb}
\usepackage{xargs}
\usepackage[pdftex,dvipsnames,table]{xcolor}
\usepackage{todonotes}
\usepackage{soul}
\usepackage{booktabs}
\usepackage[hyphens]{url}
\graphicspath{{./figs/}}

\DeclareTextFontCommand{\hyphenatedtexttt}{\ttfamily\hyphenchar\font=45\relax}

\newlength{\vcolumnwidth}
\newcommand{\mc}[1]{\mathcal{#1}}
\newcommand{\sysname}{\textsc{Cimplifier}}
\newcommandx{\note}[2][1=]{\todo[linecolor=OliveGreen,backgroundcolor=OliveGreen!25,bordercolor=OliveGreen,#1]{#2}}

\begin{document}

\date{}

\title{Towards Least Privilege Containers with \sysname{}}

\author{
  \resizebox{\textwidth}{!}{%
    Vaibhav Rastogi\textsuperscript{*\dag}
    \enskip Drew Davidson\textsuperscript{*}
    \enskip Lorenzo De Carli\textsuperscript{*}
    \enskip Somesh Jha\textsuperscript{*}
    \enskip Patrick McDaniel\textsuperscript{\dag}
  } \\
  \textsuperscript{*}University of Wisconsin-Madison
  \qquad \textsuperscript{\dag}Pennsylvania State University \\
  \{vrastogi, davidson, lorenzo, jha\}@cs.wisc.edu \qquad mcdaniel@cse.psu.edu
}

\maketitle

\input{abstract}

\input{intro}

\input{overview}
\input{design}

\input{eval}
\input{related}

\input{discussion}

\input{conclusion}

\theendnotes

{\footnotesize \bibliographystyle{acm}
  \bibliography{main}
}

\end{document}

%% file: abstract.tex
\subsection*{Abstract}

Application containers, such as Docker containers, have recently gained
popularity as a solution for agile and seamless deployment of
applications. These light-weight virtualization environments run
applications that are packed together with their resources and
configuration information, and thus can be deployed across various
software platforms.  However, these software ecosystems are not
conducive to the true and tried security principles of privilege
separation (PS) and principle of least privilege (PLP). We propose
algorithms and a tool \sysname{}, which address these concerns in
the context of containers. Specifically, given a container our tool
partitions them into simpler containers, which are only provided
enough resources to perform their functionality.  As part our solution,
we develop techniques for analyzing resource usage, for performing
partitioning, and gluing the containers together to preserve
functionality. Our evaluation on real-world containers demonstrates
that \sysname{} can preserve the original functionality, leads to
reduction in image size of 58-95\%, and processes even large containers
in under thirty seconds.

%% file: intro.tex
\section{Introduction}

In the past few years, the information technology (IT) industry has adopted
containers for addressing issues related to agile deployment of
software. {\em Containers} are light-weight virtualization environments
to ``contain'' applications and provide desirable properties,  such as an
application-specific configuration, isolation, and resource
provisioning. With recent projects, such as
Docker~\cite{docker},
containers have become  the holy grail for devops: an easy and
widely-supported specification of an application and its environment that can be
deployed across various software platforms.
Indeed, surveys and analytics indicate undeniably high rates of
Docker adoption~\cite{stackengine15,datadog15}.

As the industry begins to adopt containers as
the mechanism for deploying complex applications, techniques to secure
and harden these applications must be developed.  One traditional
principle in security is that of {\em principle of least privilege
  (PLP)}~\cite{ss75,kef+05}, which dictates that any module (an
application, process, etc.) should be allowed only privileges
that are necessary to perform its functionality. However, the current
container ecosystem does not promote this important security tenet
of PLP.  While an application inside a container can access resources
only within the container only, a container typically builds on other
images, which still includes all sorts of utilities, programs, and
files that may not be required by the application's
functionality. These redundant resources are available to the
application and in an event of a compromise only serve to increase the
possibility of further harm. Frequently, a container packs a complex
application consisting of multiple application components, which goes
against the principle of {\em privilege separation (PS)}~\cite{pfh03}:
separating modules with different privileges so that a compromise of
one module limits the attacker to only a subset of privileges.
PLP and PS dictate that a container should {\it run only one simple
application and should pack only as much resources as needed to
fulfill its functionality requirement}. 

For example, consider a wiki installation, which runs a web server and
stores the documents in a database server. If both the web server and
the database server have free access to each other's resources, a
compromise of one component could escalate to the compromise of the
other.  Furthermore, if the database server does not need
\texttt{netcat} for its operations, then \texttt{netcat} should not be
available to the database server. The wiki example is further
elaborated in Section~\ref{sec:problem} and will serve as the running
example for our paper.

This paper presents the design and implementation of \sysname{}
(pronounced {\em simplifier}) to address these problems. Given a
container, we first use dynamic analysis to understand how resources
are used. We then partition the container based on very intuitive and
simple user-defined policies into multiple containers, each running
one or more application components at the granularity of individual
executables (i.e., currently, we do not partition an executable), and
pack just the resources needed to execute these components. Finally,
we provide techniques for an application component residing in one
container to be able to invoke another component residing in another
container, so that together these containers provide the same
functionality as the original container. \sysname{} does not need
application source code nor does it depend on applications using a
particular language or runtime stack (e.g, Java VM) and hence can
handle a wide class of containers. Our work resembles existing work on
classic privilege separation~\cite{bs04, cla+07, zznm01}. However,
our target context (i.e., containers) poses unique opportunities and
challenges, which makes our work significantly different from the
above-mentioned existing work. We elaborate on these differences in
our section on related work.

We evaluated our \sysname{} prototype on several real-world
containers, ranging from simple web servers and database engines to
complex applications like a wiki, a blogging engine, and a
log-analysis stack. Our evaluation demonstrates that \sysname{} is
effective in creating functional partitions guided by simple and
intuitive user-defined policies and reducing container size by
58--95\%.  Our tool also has low execution overhead (e.g., on our
largest example it runs in under {\it thirty seconds}.)

Our contributions can be summarized as follows: 
\begin{itemize}
  \item {\em Resource identification.} We develop techniques based on system
    call logs to analyze the usage of resources and associate them with various
    executables in the application container being analyzed.
    
  \item {\em Container partitioning.} We devise an algorithm for partitioning
    a container (based on a simple user-defined policy) and for associating resources
    with the components of a partition. 
    
  \item {\em Remote process execution.} We introduce remote process
    execution (RPE) as a mechanism for gluing components.  Our mechanism
    allows a process running in one container to transparently execute
    a program in another container without relaxing the separation
    boundaries provided by containers.
    
  \item {\em System prototype.} We implemented the above techniques in a
    prototype implementation called \sysname{}, which is an end-to-end system
    for container partitioning and resource slimming. Our tool takes as input
    a container, system call logs, and a user policy  and outputs partitioned
    containers, each packing only resources needed for functionality.
\end{itemize}

The rest of this paper is organized as follows:
Section~\ref{sec:overview} provides the requisite background, problem
definition, and an overview of our approach.
Our methodology is discussed in
Section~\ref{sec:design}. We present evaluation results in
Section~\ref{sec:eval}. Related work is discussed in
Section~\ref{sec:related}, followed by a discussion on limitations and
future work in Section~\ref{sec:discussion}. We conclude in
Section~\ref{sec:conclusion}.

%% file: overview.tex
\section{Overview}\label{sec:overview}
This section provides the relevant background, the problem statement,
and issues specific to the container ecosystem. We also provide
a brief overview of our solution.

\subsection{Background}
Containers are potentially-isolated {\it user-space instances} that share the same
kernel. Mechanisms exist in several operating systems for providing multiple
isolated user-space instances. Solaris Zones and FreeBSD Jails are some
well-known examples. The Linux kernel implements {\em namespaces} to provide
user-space instantiations.

A namespace is an abstraction around a global resource giving the
processes within the namespace an illusion of having an isolated
instance of the resource, i.e., any operations on the resource are
visible to the processes within the namespace only. Six kinds of
namespaces are defined in Linux: IPC (inter-process communication),
network, mount, PID (process identifier), user, and UTS (UTS stands
for Unix timesharing system and the namespace allows for separation of
hostnames, for example). Processes with appropriate Linux
capabilities\endnote{Linux capabilities are units of privileges that
  can be independently enabled or disabled. It is thus possible for a
  superuser process to have only a limited set of privileges.} can
spawn new namespaces through the {\tt clone} or {\tt unshare} system
calls.

Container implementations on Linux, such as
LXC~\cite{lxc} and Docker, employ the
namespaces feature to provide a self-contained isolated
environment. This isolation is based on the fact that resources that
do not have a name in a namespace cannot be accessed from within
that namespace. For example, a process outside a given PID namespace cannot
be specified by a PID in that namespace. In addition, container
implementations use {\em cgroups}, another Linux kernel feature
allowing for resource limiting, prioritization, and accounting.
Together cgroups and namespace provide mechanisms to build container
security (cgroups provides protection against denial-of-service (DOS)
while namespace isolation provides basic sandboxing). Finally, Linux
capabilities and mandatory access control (MAC) systems, such as
SELinux or AppArmor, are often used to
harden the basic namespace-based sandboxing~\cite{dockersec}.

Besides the implementation of a container itself using primitives
provided by the kernel, projects such as Docker and the Open Container
Project also made strides in developing specifications and tools
for implementing and deploying containers. For example, the files
necessary for running applications (the application code, libraries,
operating system middleware services, and resources) packed in one or
more archives together with the necessary metadata constitute a {\em
  container image}. The image metadata includes various configuration
parameters, such as environment variables, to be supplied to a running
container, and network ports that should be exposed to the host.

Systems like Docker are designed particularly with the idea of
deploying applications and hence are meant to run {\em application
  containers} as opposed to {\em OS containers}. That is, they offer
mechanisms more suited to run simple applications (such as a HTTP
server) rather than a full operating system with a slew of services
and processes. In this regard, a container may be viewed as an
application packed together with all the necessary resources (such as
files) executing in an appropriate environment.  The focus of this
work is such application containers, henceforth referred to as simply
{\em containers}, and we demonstrate our methodology in the context of
Docker and its ecosystem, although our techniques are applicable in
other contexts.

\subsection{Problem Statement}
\label{sec:problem}

As mentioned earlier, an application container packs a given
application together with its resources and configuration so that the
application can execute independently on a host as an atomic
unit. Based on the principle of least privilege, an ideal container
should satisfy two requirements: {\sf (A): Slim} it should pack no
more resources than what its functionality needs, {\sf (B):
  Separation} it should execute only one simple application.
Requirement A is motivated by reducing the attack surface and
requirement B is motivated by ease of hardening (simple applications
are easy to harden using systems such as SELinux and
AppArmor). Moreover, a simple application is also more auditable
(e.g., using existing static-analysis tools, such as Coverity and
Fortify) than a complex one. Furthermore, running one simple
application per container aligns with the microservices philosophy
whereby complex applications are composed of independently running
simple applications that are easier to manage and deploy.

These requirements mandate that complex application containers be
decomposed or partitioned into minimal simple containers. This paper
is a step towards automatically performing this task. Next, we provide
a running example, which will be used at various points in the paper,
and then state a general form of the container partitioning problem.

\begin{figure}
  \centering
  \includegraphics[width=0.8\vcolumnwidth]{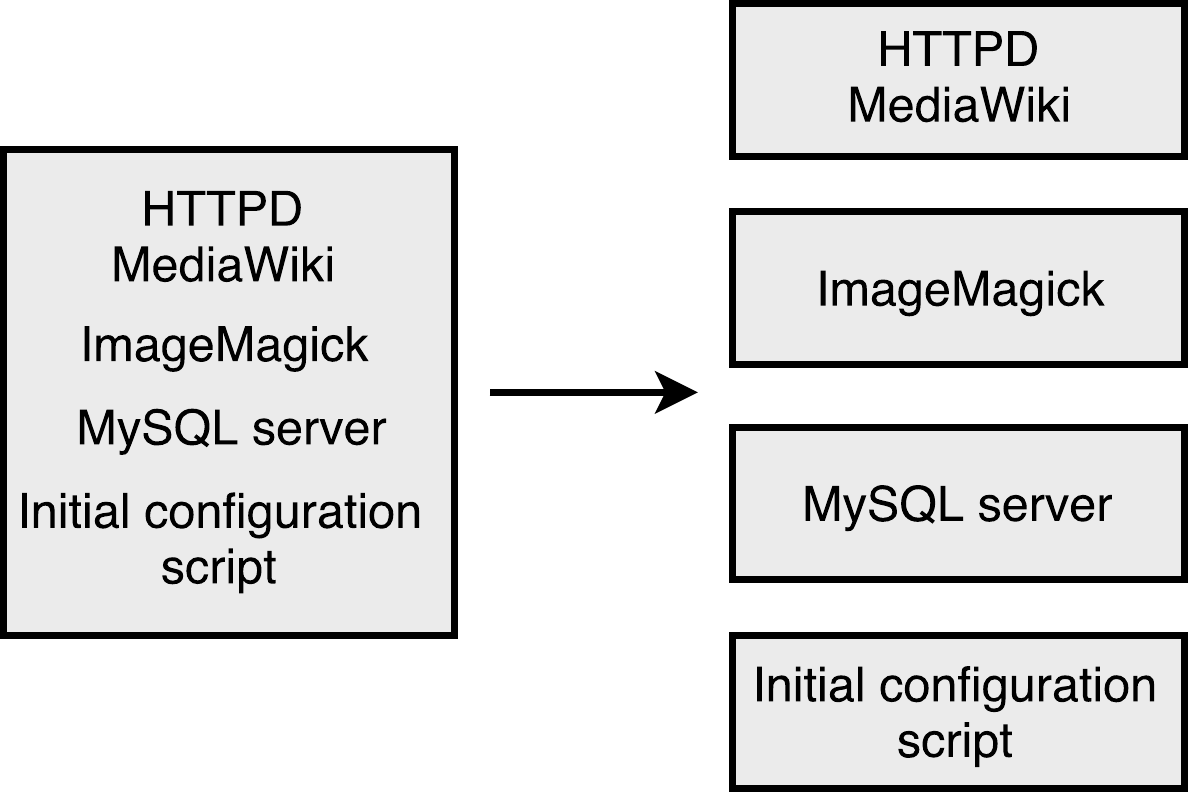}
  \caption{Running example. The figure shows the original container with its
  application components, and a desired partitioning of the components into
different containers.}
  \label{fig:mediawiki}
\end{figure}

\paragraph{Running example.}
We consider a popular container from the Docker Hub (the official repository for
container images) called \texttt{appcontainers/mediawiki}. As the name suggests,
this container provides, MediaWiki, a popular wiki application~\cite{mediawiki}. The container
may be used in different modes; we will discuss only the stand alone mode, which
allows running the entire application from just this container. The container
image has Apache HTTPD and MySQL server installed. At startup, it
performs configuration to set up the MediaWiki PHP source code for
use with HTTPD, set up a TLS certificate using OpenSSL for serving HTTPS
connections, and start and configure MySQL and HTTPD servers.

Since HTTPD and MySQL server are separate entities, and so to maximize
isolation between them, we would like them to execute in different
containers. The MediaWiki PHP code also spawns ImageMagick to convert
uploaded images to different sizes. Since uploaded images may be
maliciously crafted, we would like to separate their processing by
ImageMagick from the rest of the application, i.e., ImageMagick should
run in a separate container. The transformation we would like to
achieve is depicted at the right of Figure~\ref{fig:mediawiki}.

\paragraph{Container partitioning problem.}
A container $C$ is a set of executables $E=\{ e_1,\cdots,e_n \}$.
Let $\mc{R}(C)$ be the set of resources used by the container $C$ (
the resources could be files, socket objects, etc.).
Moreover, there are two kinds of user constraints {\it positive}
and {\it negative} (the positive constraints $UC_{+} \subseteq E \times E$
and the negative constraints $UC_{-} \subseteq E \times E$). Intuitively,
if $(e_i,e_j) \in UC_{+}$, then executables $e_i$ and $e_j$ should be
put in the same container.
Similarly, if $(e_i,e_j) \in UC_{-}$, then executables $e_i$ and $e_j$ should be
{\it not} be put in the same container.

Now the {\it container partitioning problem} can be defined as
follows: Given a container $C$, $UC_{+}$ and $UC_{-}$, find a set of
containers $\{ C_1,\cdots,C_k \}$ (where $C_i \subseteq E$, $C_i \cap
C_j = \emptyset$ if $i \not= j$, and $\cup_{i=1}^k C_i \; = \; E$ (or
in other words $\{ C_1,\cdots,C_k \}$ is a partition of  $E$).

The set of resources needed by a container $C_i$ (denoted by
$\mc{R}(C_i)$) is implicitly defined. Note that additional constraints
(e.g. minimizing the number and size of the containers) can be easily
added to the problem definition.  Our solution strategy takes into
account these additional constraints. For example, one constraint
might be to find a partition $\{ C_1,\cdots,C_k \}$ that minimizes
$\max \{ | \mc{R}(C_1) |, \cdots , | \mc{R}(C_k) | \}$. To perform the
original functionality, the containers in $\{ C_1,\cdots,C_k \}$ must
cooperate with each other, such as through network or inter-process
communication or even shared, regular files.

The above problem entails reducing the set of required resources. In fact, a
good algorithm will attempt to minimize the resource set of each container so
that the container does not have any redundant or unused resources. Such
containers are referred to as {\em slim containers}. In the Docker ecosystem, it is
common to have simple images built on top of ``fat'' operating system images (Docker
has a concept of layered images, where a container image can be defined by a few
changes over another image), so that running even a simple application container
may take hundreds or thousands of megabytes of disk space. For example, the
official Python image, which one might consider as the base for a simple Python
application, is over 675 MB! Needless to say, much of the file content in this
image is never used. With such huge image sizes being a rule rather than an
exception in the Docker ecosystem, numerous articles and blog posts have been
written about slim containers~\cite{d15a,d15b, h15}. Our algorithm
tries to reduce the size of the resources associated with a container $C_i$,
i.e., we {\em
achieve slimming automatically}.

\paragraph{Other benefits.}
We also see new opportunities in a good solution for container partitioning.
First, it may enable an altogether new workflow, whereby a system administrator
starts off by installing applications in one monolithic container, as is done in
a traditional operating system, and then reap benefits of automatic container
partitioning.  In such a workflow, the system administrator need not be
concerned about how different application components need to connect with each other
(as she would be when manually preparing containers for each of them). Second,
container partitioning could be an enabling precursor to automatic
multi-location deployment of complex applications: partitions of containers
could be distributed across multiple locations (physical servers or geographic
locations). This could be done for reasons like load balancing or privacy 
compliance (such as, certain data resources should remain within an enterprise
or be encrypted).

\begin{figure*}
  \centering
  \includegraphics[width=\textwidth]{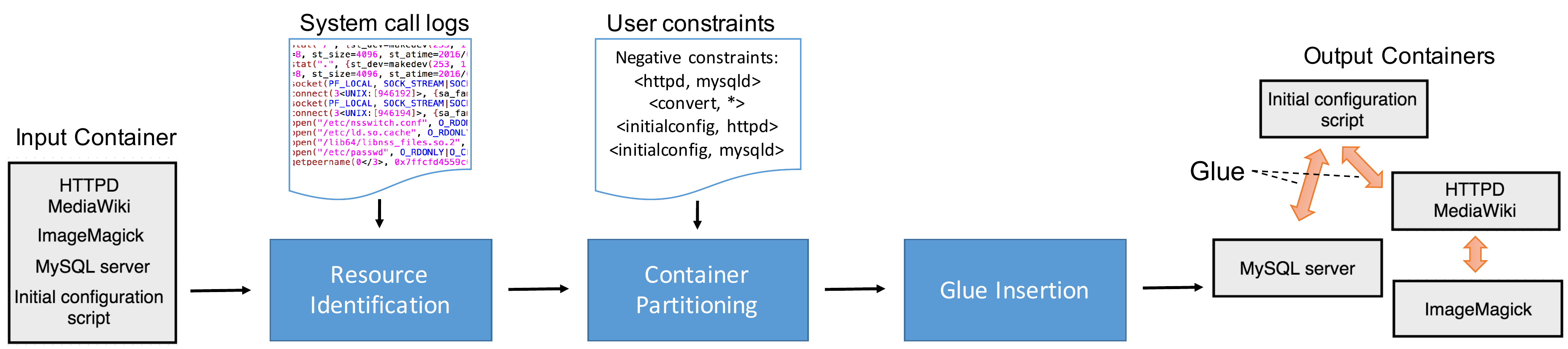}
  \vspace{-0.15in}
  \caption{Architecture overview. We analyze system calls from model runs of the
    input container to identify resources. The application together with these
    resources is then partitioned across several containers guided by a
    user-defined policy. These containers function together through remote
    process execution, which acts as a glue among them.}
  \label{fig:arch}
\end{figure*}

\subsection{Our Approach}\label{sec:approach}
Given a container, our work partitions it at the level of application
executables so that an executable binary is one atomic unit that can be placed
in a partition. Partitioning at granularities finer than executables is not
within the scope of the current work but may be achieved by combining our work
with previous work on program partitioning and privilege
separation~\cite{bs04,clm+09,cmam12}.

Container partitioning poses three technical challenges {\sf (A)} How do we
identify which resources are necessary for a container? {\sf (B)}
How do we determine container partitions and associate  resources with them? {\sf (C)}  How
do we glue the container partitions so that together they provide the same
functionality as the original container?

Our approach utilizes dynamic analysis to gather information about the
containerized application's behavior and functionality. We collect
detailed logs from executions of a given container. We use these logs
to construct resource sets for different component executables. Based
on flexible, pluggable policies, container partitions are
determined. The resulting containers are populated with the resources
needed for correct functioning of the executables. Container
mechanisms themselves provide for complete isolation among
containers. Based on the resource sets identified, we relax this
isolation to share some resources across containers on an as-needed
basis. Finally, we introduce a new primitive called {\em remote
  process execution} to glue different containers. It relies on the
availability of a shared kernel to allow a process to transparently
execute a program in a different container. The approach is
schematically depicted in Figure~\ref{fig:arch}.

While our approach uses dynamic analysis, partitioning may also be
possible through a static analysis. Both the approaches have their
advantages and disadvantages. In dynamic analysis, resource set
identification may not be accurate if code coverage during container
executions is not complete. Static analysis does not suffer from this
limitation but faces significant challenges in our context: in typical
containers that we studied application components are written in
several languages (e.g., shell script, PHP, and compiled C and C++),
the application is strewn across multiple shared object libraries and
executables, and the content of the environment variables and
configuration files dictate an application's runtime behavior. Our
dynamic analysis on the other hand stays on the simple, well-defined
system call interface and is more manageable. Solutions combining
static and dynamic analyses to incorporate the advantages of both will
be interesting to investigate in the future.

%% file: design.tex
\section{Methodology}
\label{sec:design}

Our algorithm has three main steps. {\em Resource identification:} In
this step we identify the accesses of different files, IPC, and
network objects by various processes in executions of the original
container. {\em Partitioning:} This step utilizes the user policies
and some other information to partition the original container into
several containers.  {\em Gluing:} In this step we ``glue'' the
containers together to preserve functionality: we introduce {\em
  remote process execution} as the mechanism to glue containers. This
section details these three steps and describes a few interesting
details of our system.

\subsection{Resource Identification}
\label{sec:rsrcidfy}

Resource identification is the first step in our methodology that
enables association of various resources, such as file, IPC and
network objects, to the subjects, i.e., the entities that act upon
them. The next steps that perform the actual partitioning use this
information to decide the resources associated with each
container. For collecting this information, the system call interface
serves our purpose well because it allows
complete access to the information exchange happening between the
user-space processes and the kernel. Actual resource access,
management, and so on happens inside the kernel, and so a process must
make system calls to the kernel to perform any action related to
resources.  There are several options for performing system call
logging, which we discuss in Section~\ref{sec:impl}. Our methodology
just needs system call logs and does not depend on any specific
logging infrastructure.

\paragraph{Analyzing system call logs.}
Let a {\em system call event} be defined as a tuple $E = \langle i, c,
\rho \rangle$, where $i$ represents the thread ID of the caller, $c$
is the name of the system call (e.g., \texttt{open} or
\texttt{rename}), $\rho$ is a sequence of parameters to the system
call (the last element of this sequence is the return value of the
system call).  Each system call has an associated type which determines how to
interpret a parameter (e.g., whether a parameter should be interpreted
as a pathname or as an integer return code). A system call {\it log}
is simply a sequence of system calls and an {\it execution log} is a
set of logs. Given a log $\sigma \; = \; \langle E_1,E_2, \cdots , E_m
\rangle$ we define $\Gamma_j$ as the state of the system after the
sequence of system calls $E_1,E_2,\cdots,E_j$ is executed (we assume an initial
state $\Gamma_0$).  Note that the event $E_k$ is executed in the state
$\Gamma_{k-1}$. Using the semantics of system calls we can define for
each tuple of a system call $E$ and state $\Gamma$ a tuple $(R,W)$,
where $R$ and $W$ are resources that system call $E$ reads from and
writes to when executed in state $\Gamma$.  We call this function
$\mathit{rsrc}$ (i.e. $\mathit{rsrc} (E,\Gamma) = (R,W)$). Note
that this function can be ``lifted'' to logs and execution logs via
the standard collecting semantics. For example, for a log $L=\langle
E_1,E_2,\cdots,E_m\rangle$, we have that $\mathit{rsrc}(L,\Gamma)$ is
equal to
\[
(\cup_{E_j \in L} \mathit{rsrc}_1 (E_j,\Gamma_{j-1}),\cup_{E_j \in L} \mathit{rsrc}_2 (E_j,\Gamma_{j-1}))
\]
In the equation given above $\Gamma_j$ is the state reached from $\Gamma$ after
executing the sequence of system calls $E_1,\cdots,E_j$ and $ \mathit{rsrc}_i$
for $i \in \{1,2 \}$ is the $i$-th component of the tuple. Given an
execution log $EL = \{ L_1 , \cdots , L_n \}$ and a state $\Gamma$
we define $\mathit{rsrc}(EL,\Gamma)$ to be 
\[
(\cup_{L \in EL} \mathit{rsrc}_1 (L,\Gamma),\cup_{L \in EL} \mathit{rsrc}_2 (L,\Gamma))
\]
These sets play a crucial role in deciding which resources are
exclusively associated with a container and which resources are
shared between containers. Next, we describe how different kinds of resources,
such as files, IPC and network, are handled in the above framework.

\paragraph{Files.} Files are handled through numerous system calls but all of
them map neatly to the above abstractions. Intuitively, a file that
should exist for the success of the call is placed in the read
set. Creation, such as by opening a previously non-existing file
or by renaming files, modification such as opening in append
mode, and modification of metadata such as by using \texttt{chmod} all
result in adding the file to the write set.  Files and other resources
referred to by file descriptors in the system call arguments are not
added to these resource sets. The reason is that a file descriptor
already had a resource associated with it when it was created and it
was at the time of creation, such as through the \texttt{open} system
call, that the resource was added to the resource set. File
descriptors are capabilities and may be passed across processes. A
process may be able to perform system calls on a file descriptor that
was passed to it from elsewhere even if the associated resource is not
directly available to it. Thus it is not necessary to make the
resource available in the container running this process.

\paragraph{Inter-process communication.} There are several inter-process
communication (IPC) options available; we support an important subset
of them. Any IPC that happens without naming a resource, such as that
through channels created by \texttt{pipe} or \texttt{socketpair}
system calls, is implicitly supported in \sysname{}. Such IPC
typically depends on file descriptor inheritance, support for which is
described in Section~{\ref{sec:rpi}}.  Named pipes or FIFOs are one
IPC mechanism that use named resources. Named pipes exist as file
system nodes; one process opens the pipe for reading and another for
writing. This establishes a communication channel between the two
processes.  Since named pipes reside on the file system and are opened
in the same way as regular files are, they are handled in the same way
as files. Unix domain sockets are another IPC mechanism. When a
process \texttt{bind}s to a socket it specifies the socket address,
creating a node named by that address on the file system (unless it is
an abstract address). Another process may \texttt{connect} to this
socket by specifying the socket address. For the success of the
\texttt{bind} call, the parent directory of the socket should exist,
making it a candidate to be placed in the read set. The socket address
file itself is considered as being modified by both the binding and
the connecting processes and is hence placed in the write set.  There
are other IPC mechanisms such as message queues and semaphores that we
do not currently support. We have however not found them to be used in
any of the containers we evaluated and hence did not face difficulties
in partitioning due to this reason. Lack of some IPC mechanisms is not
a fundamental limitation of our approach and can be addressed by
further engineering effort.

\paragraph{Network communication.} Network communication is sometimes used
within a container such as by a web server to connect to a backend
database server. Unlike Unix domain sockets, in the case of network
sockets, the socket address is not a file node and a \texttt{bind} or
\texttt{connect} call is not needed for connectionless sockets.
Nonetheless, for connectionless sockets, the socket address is
provided in the \texttt{recv}* and \texttt{send}* system calls. This
socket address is placed in write set.

\subsection{Partitioning}

Let $E = \{ e_1, e_2, \cdots, e_n \}$ be the set of executables in a
container.  On Linux, any run of an executable starts with the
\texttt{execve} system call (the other \texttt{exec} family functions,
and \texttt{system} and \texttt{popen} functions ultimately make this
system call).  Using the semantics of \texttt{execve} we can associate
an executable $e_i \in E$ to each system call in the execution log
(i.e., the system call originated from a function in that executable).
Using this information and the result of the resource identification
step, we can associate a set of resources read and written by an
executable. We next describe our partitioning algorithm.

\noindent
{\bf Our partitioning algorithm.}  Let $E={e_1,\cdots,e_n}$ be the set
of executables in a container. With each executable $e_i$ we
associate a tuple $ (R(e_i),W(e_i))$, where $R(e_i)$ and $W(e_i)$ are
the resources read and written by $e_i$, respectively. Note that given
a subset of executables $E' \subseteq E$, the tuple associated with it
is $(\cup_{e \in E'} R(e), \cup_{e \in E'} W(e) )$.

Let $G=(E,X)$ be a directed graph whose set of vertices is the set of
executables $E$ and the set of edges $X \subseteq E \times E$ is such
that $(e_i,e_j) \in X$ iff $e_i$ calls $e_j$ (some function in
executable $e_i$ executes $e_j$). In other
words, $G$ is the {\it call graph} at the executable level. Currently,
we construct $G$ dynamically by inspecting the execution logs, but it
could be easily constructed statically.

Let $UC_{+} \subseteq E \times E$ and $UC_{-} \subseteq E \times E$ be
the positive and negative constraints provided by the user. For each
executable $e_i$, let $C(e_i)$ be the current index of the container
in the partition. Our algorithm works in two steps as follows:

\noindent
{\bf Initial partition based on user constraints:} Each executable is
in a single container, i.e.  $C(e_i) \; = \; i$. For each, constraint
$(e_k,e_j) \in UC_{+}$ we merge the containers that have $e_k$ and
$e_i$ (i.e., we merge the containers $C(e_i)$ and $C (e_j)$ into one
container. Note that this can be performed by using a simple
re-indexing).  However, we {\it do not} perform a merge corresponding
to a constraint in $(e_i,e_j) \in UC_{+}$ if it will violate any
negative constraint $UC_{-}$ (i.e. there is $(e_i,e_j) \in UC_{-}$
such that $e_i$ and $e_j$ will be in the same container after the
merge). We keep merging containers using this rule, until we reach a
fix-point (i.e., the partition induced by $C(\cdot)$ does not change).

\noindent
{\bf Updating the partition based on the call graph:} Intuitively, if
$(e_i,e_j) \in X$ (some function in $e_i$ calls a function in $e_j$),
then $e_i$ and $e_j$ should be in the same partition as long as the
given constraints are not violated. For each edge $(e_i,e_j) \in X$ we
merge the containers $C(e_i)$ and $C(e_j)$ as long as any negative
constraint in $UC_{-}$ is not violated. We keep merging containers
based on the call graph until we reach a fix point.

In our partitioning algorithm we have experimented with three types of
user defined policies, which can be easily specified using our formalism.

\noindent
{\it All-one-context.} This policy places all executables into a single
    container. Thus, it does not perform any container partitioning. However,
    since only the resources accessed during test run are placed in the
    container, this policy is tantamount to container slimming. Thus container
    slimming is just the simplest configuration of \sysname{}. In our formalism,
    this corresponds to $UC_{+} = E \times E$ and $UC_{-} = \emptyset$.

\noindent    
{\it One-one-context.} This policy places each executable into a
separate container so that no two executables share containers. While this
policy is useful for testing \sysname{}, it is not practical for
reasonably complex containers that may involve tens or even hundreds
of executables. (The container of our running example uses $49$
different executables, including simple utilities like \texttt{cat}
and \texttt{tail} as well as related executables, which together can
be considered as one component, like HTTPD.  Putting each executable
in a separate context is unnecessary in these cases.) This policy
corresponds to $UC_{+} = \emptyset$ and $UC_{-}$ contains all pairs
$(e_i,e_j)$ such that $i \not= j$.

\noindent
{\it Disjoint-subsets-context.} In this policy, the user specifies
disjoint subsets of executables, not necessarily covering all the
executables. The subsets correspond to different containers. That is,
a container corresponding to a given subset contains executables in
that subset but in no other subset. Some executables may not have been
specified and are considered don't cares; they can be placed in any
container. This policy is practically useful. In our running example,
one can specify the HTTPD-related executables in one subset,
MySQL-related ones in another and ImageMagick-related ones in
another. Other executables such as \texttt{cat}, \texttt{mv}, and
\texttt{chmod} are small and may not be considered sensitive to
warrant separate containers and hence can be placed anywhere as needed
to optimize performance.

Apart from the above policies, it is possible to implement more
intricate policies as well. For example, consider two programs both
invoking \texttt{openssl} to configure cryptographic keys. If the user
would prefer that the two invocations of \texttt{openssl} from the two
programs happen in separate contexts to prevent any interference and
protect cryptographic keys, it is possible to define a policy that
does so. Such a policy will separate the two \texttt{openssl} contexts
by the invoking contexts: if the context of first program invoked
\texttt{openssl}, then run it in one container else run it in the
other container. The above is an example of a policy based on
execution-history~\cite{Abadi2003}. Our policy language can be
readily extended to support history-based policies. 

\paragraph{Resource placement.}
To associate resources with containers (we call this resource placement), we
begin with resources read by its executables.
There are some tricky issues that arise here.
For example, before
placement, this ``read'' set must be extended to cover all dependencies. In
particular, if a file indicated by a given path is placed in a
container, we must ensure that all the directory components in the
path leading up to the file are also placed in the container. In
addition, while placing resources, we make sure that their metadata
(such as permissions, ownership, and modification times) is preserved.

A resource may need to be placed in multiple containers. In such a
case, the nature of resource access determines how the placement is
performed. If a resource is read-only, the resource can be safely
duplicated: each container gets its own copy of the resource. If a
resource is modified or created by one of the containers, it should be
shared. Docker natively provides the capability of mounting host files
at given mount points inside a container. Such mounts are called {\em
  volumes}. Volumes may be shared between containers, i.e., one host
file may be simultaneously mounted in multiple containers. For file
resources as well as named pipes and Unix sockets, we use shared
volumes for shared resources. Note that in case a resource is created
by a container at runtime and is used by another container, the parent
directory of the resource must be shared. This is because Docker
volumes may only be mounted when a container is started and Linux does
not allow mounting of non-existent files. Our solution does not
differentiate between created files and modified files and hence, we
always mount parent directories when resources are shared. Sharing
parent directories can result in over-sharing of resources; this
compromise appears necessary, however, as volumes appear to be the
only way to share resources among containers.

For sharing network resources, we need to match socket addresses from
the write set. This match is not an exact string match but made according to
the socket address specification semantics. For example, the bound,
listening TCP/IP address $\langle 0.0.0.0:3306 \rangle$ and the
connecting TCP/IP address $\langle 127.0.0.1:3306 \rangle$
match. Based on such socket address matches, we determine the
containers that need to communicate over the network. Since the client
(connecting or sending process) may try to connect at the
localhost/loopback address (as it may have been configured in the
original container) but now the destination address is the hostname/IP
address of the other container, we need to run a port forwarder that
forwards connections and messages from a localhost address to the
remote container address. This has some overhead and realizing that
sharing the network namespace may not introduce significant security
issues, our current implementation takes the option of sharing the
network namespace across the created containers.

\subsection{Gluing}
\label{sec:rpi}

The remaining technical challenge in obtaining a functional
partitioned containers is to ``glue'' them together to maintain the
original functionality.  Our technique to handle this challenge is
{\em remote process execution (RPE).}  RPE transparently allows a
process to execute an executable in a container different from the one
in which it resides. By {\em transparency}, we mean that neither
of the two executables (caller and the callee) need to be
modified or made aware of remote process execution.

Returning back to our running example, MediaWiki uses ImageMagick to
create thumbnails of uploaded images. Since Mediawiki's PHP code runs
in the HTTPD container and ImageMagick, corresponding to the
\texttt{convert} executable, runs in a separate container, simply
executing \texttt{convert} from PHP code will fail because the
executable file \texttt{convert} does not exist in the HTTPD
container. We need a technique that allows the PHP code to execute
\texttt{convert} and yet the actual execution of \texttt{convert}
should happen along with its resources in the ImageMagick
container. In the same running example, the need to execute processes
in other containers arises elsewhere as well: the startup script
invokes executables in HTTPD and MySQL containers to start the HTTPD
and MySQL servers, respectively. In general, RPE serves as the
fundamental glue primitive among partitioned containers and is crucial
for preserving functionality.

\begin{figure}
  \centering
  \includegraphics[width=\vcolumnwidth]{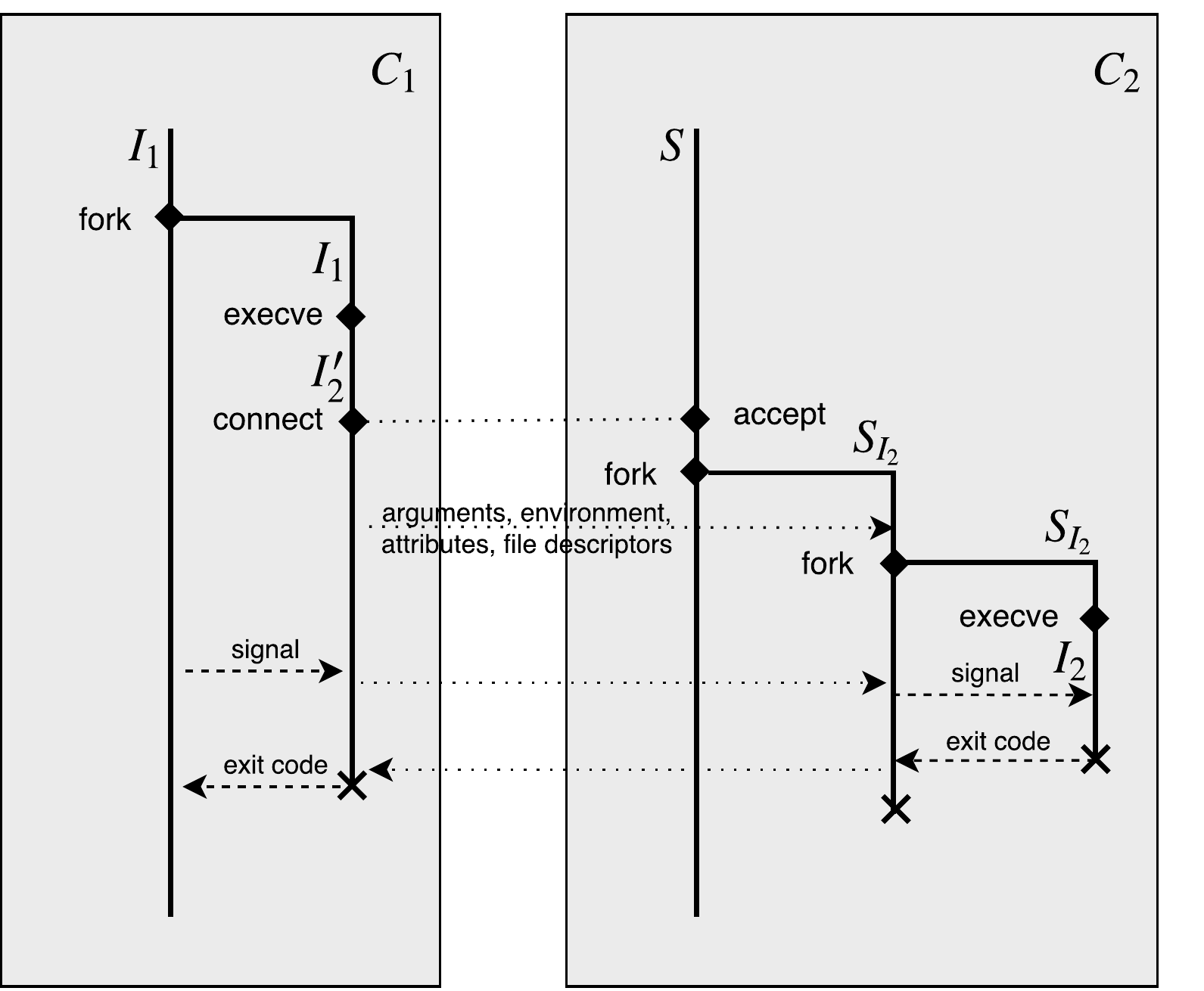}
  \caption{Remote process execution (RPE). The executable $I_1$ in container $C_1$
  transparently executes $I_2$ in container $C_2$ with our RPE mechanism. Dark
lines indicate processes; time progresses downwards. Dotted lines indicate socket
communication and dashed lines indicate signal and exit code propagation.}
  \label{fig:rpe}
\end{figure}

Our solution works as follows: consider the scenario in which
an executable image $I_1$ in container $C_1$ needs to execute
another executable image $I_2$ that actually resides in container
$C_2$. Our solution is to place a stub $I'_2$ corresponding to the
actual executable $I_2$ in container $C_1$. We also run a ``server''
in $C_2$ to accept RPE requests. When $I_1$ executes $I_2$, it is
actually $I'_2$ that is executed ($I'_2$ is on the same path as
$I_2$). $I'_2$ marshalls the command arguments, its environment, and
some process attributes and sends an RPE request to the server running
on $C_2$, which then executes $I_2$ on behalf of $I_1$.  This scenario
is described in Figure~\ref{fig:rpe}. The concept
of RPE strongly resembles remote procedure calls (RPCs) where a process
invokes a stub procedure, which marshalls the arguments supplied and
sends them to the remote server, which unmarshalls the arguments and
calls the actual procedure with those arguments. The key difference in
our context is that instead of shipping just the arguments, we also
need to ship the process environment as well as low-level process
attributes. Our key insight is that {\it the local and remote processes
  share a kernel and thus appropriate shipping of process
  attributes and file descriptors can make RPE transparent to the
  participating programs}. RPCs cannot assume a
shared kernel and therefore cannot provide as rich features and
transparency as we can. When the process ends, we ship the return code
of the process to the executing process. In addition, we provide for
asynchronous signaling (i.e., passing POSIX signals) between
processes. All these aspects require a detailed understanding of the
specification of a process. To highlight the difference between RPE
and RPC we describe one mechanism (i.e., how attributes are replicated
at start). Due to lack of space we will not describe other details.

\paragraph{Attribute replication at start.}
$I'_2$ obtains the attributes to be replicated through the relevant
system calls while the server side sets those attributes through
another set of relevant system calls just before executing $I_2$. Some
of the attributes, such as the user and group IDs, may be set only by
a privileged user. Therefore, $S$ runs as root. The arguments and
environment arrays are readily available to $I'_2$, and thus easily
copied over, and the remote side can execute $I_2$ supplying the same
arrays. The replication of file descriptors (henceforth called fds)
needs further discussion. $I'_2$ may discover its open fds to be
replicated by examining the entries of \texttt{/proc/self/fd} and
filtering out the fds it itself opened (particularly, the one opened
to read this directory and possibly also the socket fd connected to
the server). Although fds are \texttt{int}s, they have a special
meaning associated to them by the kernel and thus have to be
transferred to other processes using the functionality provided by the
kernel (this highlights one of the key differences with RPC). One
common method of transferring fds is through ancillary messages
feature of Unix domain sockets, which is why our client and server use
these sockets. At the server side, the received fds may not have the
same number as the originals. For example, the fd $2$ may have been
received as fd $4$. This final conversion of a descriptor to the right
number can be done by using \texttt{dup2} system call and then calling
\texttt{close} on the original descriptor. Thus, we have made sure
that if $I_2$ expects an open fd $3$, for example, from $I_1$, it will
receive that same fd number even when executing remotely. Some readers
might legitimately wonder why replication of file descriptors is
necessary. The answer is that file descriptor inheritance is behind
the functioning of numerous mundane tasks. An example will make this
clear: upon executing ``\texttt{cmd~>~file1}'' in a shell, the shell
creates a new process (by \texttt{fork}, for example); opens
\texttt{file1} and duplicates (using \texttt{dup2}) the newly opened
file descriptor to fd 1 (corresponding to \texttt{stdout}) and then
executes cmd1. A similar creation of file descriptors and duplication
happens for piping output, as in ``\texttt{cmd1 | cmd2}''. This makes
replication of file descriptors crucial for transparent
execution.

\paragraph{Generalizations.}
It is worth noting that the concept of remote process execution as
described above is actually independent of containers but we do need a
shared kernel because of the need to replicate file descriptors. Thus,
the technique could be applied to other applications where an
unprivileged or sandboxed application should execute a process in
another sandbox, for example a SELinux sandbox~\cite{w10}. It is also
worth noting that the techniques described above are not specific to
Linux also. Even though our implementation uses Linux-specific API and
constructs for concreteness, it is possible to make it portable across
most of the modern Unix platforms.  Thus our RPE mechanism should also
be applicable to container mechanisms on other operating systems, such
as FreeBSD Jails or Solaris Zones.

\subsection{Implementation}
\label{sec:impl}

We have implemented a prototype of \sysname{}. Our implementation consists of
about 2,000 lines of Python and C code, with C being used to implement RPE.

Our resource identification relies on system call logs, which can be
collected in several ways, such as by hooking \texttt{libc} functions
that wrap system calls by techniques like \texttt{LD\_PRELOAD}; by
using the \texttt{ptrace} system call, which allows a tracer process
to collect information from and control a tracee process; by using
in-kernel information collection modules, such as
SystemTap~\cite{systemtap} or SysDig~\cite{sysdig}; or by using
Red Hat's Audit system~\cite{rhaudit}.
Each has their advantages and disadvantages; we used
\texttt{ptrace}. Specifically, we use the \texttt{strace} utility,
which uses \texttt{ptrace} and is meant for system call
logging.

Our prototype makes a few deviations from the resource identification
semantics described earlier. One particularly interesting case is that of
unsuccessful system calls. The system calls may fail for variety of
reasons and finding the exact reason for failure requires emulating
the relevant kernel code in its entirety (note that \texttt{errno}
gives only a broad reason for failure). In our prototype
implementation, for most part, we ignore unsuccessful system calls as
if they never happened. If a program interprets the reason (through
\texttt{errno}, for example) for a failed system call, we may not be
able to capture this behavior. For remote process execution, we do not
currently replicate some less commonly used attributes. For example,
we do not replicate nice value and resource limits. For the containers
we have studied, such attributes remain at default value across the
container boundaries, so our experiments are not affected. A
production deployment however would need to perform a complete
replication, especially given that it is relatively straightforward to
implement these features.

%% file: eval.tex
\section{Evaluation}\label{sec:eval}

This section presents our experiments to evaluate \sysname{} and the results
thereof. Our evaluation seeks to answer the following questions.
\begin{itemize}
  \item Does \sysname{} work on real-world containers and do its output
    containers preserve the functionality of the input containers?
  \item How effective is \sysname{} at partitioning real-world containers and
    eliminating redundant resources from them?
  \item How much time does \sysname{} take to analyze the inputs and produce the
    outputs?
  \item What is the runtime overhead of the output container system
    produced by \sysname{}?
\end{itemize}
Our experiments are divided into two parts: experience with real-world,
popular containers and a microbenchmark to measure runtime performance. We
summarize the findings from our experiments and then describe the latter in
detail.
\begin{itemize}
  \item \sysname{} succeeded in processing all containers we experimented on.
    Furthermore, the resultant containers preserve the functionality that
    the input containers were tested for.
  \item \sysname{} produced the desired partitions as specified by the user
    constraints. As for redundant resource elimination, \sysname{}
    reduces container sizes by 58-95\% in our experiments.
  \item Given input containers, their system call logs, and partitioning policies,
    \sysname{} produces output containers in under 30 seconds for even the most
    complex containers we examined. It is thus fast enough for real-world use to partition
    and slim containers.
  \item The partitioned containers did not incur any observable runtime
    overhead.
    Overhead is expected only in our glue (remote process execution), which we
    measure to be 1-4 ms on realistic programs and is amortized to
    negligibility over the entire runs of the programs.
\end{itemize}

\subsection{Case Studies}
We tested \sysname{} on seven popular containers from Docker
Hub. Four of these containers were chosen from the Docker official images
(the Docker project directly provides and maintains about a hundred so-called
official container images providing popular applications), which are simple but
highly used containers. As each of these run one simple application only, we
did not expect \sysname{} to partition them but only to perform slimming
in accordance with least privilege. The other three containers are
popular community-contributed containers and run multiple application
components that should be partitioned. We used the latest versions of these
containers, as collected in January-February 2016.

Our system call logs are collected from test runs that exercise the given
container's functionality. In a workflow using \sysname{} in practice, a system
administrator would have a specific purpose in deploying an application and
would therefore provide test runs demonstrating that one purpose only.
Therefore, where multiple configurations are possible, we
exercise our containers based on the functionality we want from them.
A preprocessing run through
system call logs shows the executables used in the container; we use this output
to write our partitioning policies.

The rest of this subsection will examine each container as a case study and then
present a discussion highlighting noteworthy lessons. Owing
to space limitations, we only discuss the most relevant or interesting details
and provide a summary of the results in Table~\ref{tab:containers}.
All the experiments were conducted
on a VirtualBox virtual machine running Fedora 23 and configured with a single
CPU core and 2GB of memory. The base hardware was a 2013 MacBook Pro with 2.3
GHz Intel Core i7-4850HQ CPU and a solid state drive.

\begin{table*}
  \centering
  \small
  \rowcolors{2}{}{lightgray!20}
  \begin{tabular}{lrrrrr}
    \toprule
    Container & \# Downloads & Size & Analysis time & Result size & Size
    reduction\\
    \midrule
    nginx & 24 M & 133 MB & 5.5 s & 6 MB & 95\% \\
    redis & 18 M & 151 MB & 5.5 s & 12 MB & 92\% \\
    mongo & 7 M & 317 MB & 14.0 s & 46 MB & 85\% \\
    python & 3 M & 119 MB & 5.3 s & 30 MB & 75\% \\
    appcontainers/mediawiki & 178 K & 576 MB & 16.8 s & 244 MB & 58\% \\
    eugeneware/docker-wordpress-nginx & 26 K & 602 MB & 16.2 s & 201 MB & 67\% \\
    sebp/elk & 59 K & 985 MB & 26.1 s & 236 MB & 76\% \\
    \bottomrule
  \end{tabular}
  \caption{Containers studied. Each row specifies the container identifier on
  Docker Hub, the number of downloads (an indicator of popularity), the
container image size, the time \sysname{} took for analysis, the combined
size of all output containers, and the percentage reduction in the size. The
first four containers are simple and hence not partitioned but only slimmed. The
containers (or container systems) produced by \sysname{} are functionally
identical to the original containers.}
  \label{tab:containers}
\end{table*}

\paragraph{Nginx.}
Nginx~\cite{nginx} is a high-performance, popular web server. We chose to use our Nginx
container as a static website server and ran tests in this configuration.
Our test runs for collecting system call logs consisted of requesting both
existing and non-existing website content through a web browser. The call logs
include only the \texttt{nginx} executable and there is no expectation of
partitioning. The website root itself exists on an
external volume and files from it are not removed (this allows the new container
to serve even files that were not served in the test runs). \sysname{} reduced
the size of this container by 95\% while keeping it functional. The reduced
container does not have a shell or any of the common Unix utilities, which
helps in mitigating further exploitation in the event of a compromise.

\paragraph{Redis.}
Redis~\cite{redis} is an in-memory data structure store that find applications as
a cache, message broker, and so on.
Our tests to exercise this container consisted of running the
\texttt{redis-benchmark} utility with the container configured in
\texttt{append-only} persistence mode (in this mode, Redis logs each operation
it executes to a file). The utility sends various commands to the Redis server,
testing all Redis operations. \sysname{} slimmed the container to just 8\% of
the original while preserving functionality.

\paragraph{MongoDB.}
MongoDB~\cite{mongodb} is a NoSQL database engine. The container runs \texttt{mongod}, the
MongoDB server. Our test runs included test cases to exercise the core
functionality of \texttt{mongod}. In particular, we used \texttt{jstests/core}
integration tests from MongoDB source repository. 45 of 825 tests failed on the
original container for unknown reasons and so were excluded from test runs.
\sysname{} was able to process the container and the logs in 14 seconds
to produce a functionality-preserving container slimmed to just 15\% of the
original.

\paragraph{Python.}
Our next container provides the Python runtime. We use this container as a base
for a simple Python web application container, which we slim with
\sysname{}. Slimming should remove even files from the Python
runtime and libraries if they are not needed to run the application. To find our
candidate web application, we explored the list of websites powered by Flask, a
popular Python web application framework (the list is curated by the Flask
project), and selected the list's first open-source website, which was also
functional on the Web. We thus selected \url{www.brightonpy.org}, the
website of a Python user group. Our test runs included setting up the website
container and opening web pages in a browser. \sysname{} could process the
resulting logs to produce a container that served the website identically while
reducing the container size from 119 MB (this includes the files added to the
Python container to setup the web application) to 30 MB. We would also like to
note that our Python
container is different from other containers we experimented on as it builds on
Alpine, a Linux distribution with a focus on space efficiency (the other
containers build on more mainstream distributions like Debian and CentOS).
Alpine-based official images are a recent effort of Docker to reduce image
sizes~\cite{c16} (most official images do not have Alpine variants yet). This
particular case study shows that \sysname{} can eliminate resources even from an
already space-efficient container.

Having discussed single-application containers, we now switch to containers that
run a full stack of applications, which should be partitioned.

\paragraph{Mediawiki.}
This study considers \hyphenatedtexttt{appcontainers/mediawiki}, the container of our
running example. As shown in Figure~\ref{fig:mediawiki}, we would like to split
this container into separate Apache HTTPD, MySQL, and ImageMagick partitions
as well as an initial configuration partition (a separate partition for initial
configuration is not strictly necessary; it is policy-dependent whether a
separate partition is created).

For our test runs, we used the MediaWiki acceptance tests. The acceptance tests
do not cover some crucial functionality, such as image uploads and table
processing. We therefore wrote additional test cases using Selenium
IDE~\cite{seleniumide}. Our test deployment includes only core Mediawiki; if
Mediawiki extensions are used, test runs should be extended accordingly. 

\sysname{} was able to produce a functionality-preserving system of four
containers. As expected, the containers need to share some volumes.
The ImageMagick and HTTPD partitions share \texttt{/tmp} and
\texttt{/var/www/html}. This is necessary because Mediawiki instructs
ImageMagick to read image files from \texttt{/var/www/html} and write converted files
to \texttt{/tmp}. The HTTPD and MySQL partitions share \texttt{/var/lib/mysql}
as MySQL creates a Unix domain socket there, which Mediawiki code connects to.
The initial configuration partition shares a few volumes (recall that we enable
file sharing by volumes) with other partitions because it needs to write some
files to these partitions before starting other services.

While we are able to eliminate most of the redundant resources, our tool failed
to reduce 125 MB of Mediawiki code. The reason was that the initial configuration
recursively calls \texttt{chmod} and \texttt{chown} on the Mediawiki directory,
touching each file there, including test cases and documentation, which could be
safely removed. Unfortunately, a pure system calls-based approach, such as ours,
cannot distinguish between intentional access and inconsequential recursive
access. We also learn that slight configuration changes could result in much
better isolation. For example, HTTPD and MySQL need to share
\texttt{/var/lib/mysql} (this directory contains MySQL database files) because
of MySQL server's socket there. If however, the socket were created at a
different location, this sharing could be avoided. Our next case study shows
this.

\paragraph{Wordpress.}
We study
\hyphenatedtexttt{eugeneware/docker-wordpress-nginx}, a container that allows running
Wordpress~\cite{wordpress}, a blog engine. It contains Nginx web server, MySQL database, PHP-FPM PHP engine, and
Supervisord (a process control system for controlling Nginx, MySQL and PHP-FPM),
each of which we would like to be placed in different partitions. The Nginx
frontend server connects via a Unix socket to PHP-FPM, which runs Wordpress code
and communicates with MySQL for storage through another Unix socket. 

Our test runs included performing actions such as creating blog posts, adding
comments, changing user profiles, and signing in and out. \sysname{}
produced the desired partitions that would together be able to serve Wordpress
just like the original container did.
It is noteworthy that the configuration of this container allows
for better isolation than the previous Mediawiki one: the mysql socket is
created in \texttt{/var/run/mysql} and so \texttt{/var/lib/mysql} (the directory
holding MySQL files and databases) needs to be present in the MySQL container
only.

\paragraph{ELK.}
The Elasticsearch-Logstash-Kibana (ELK) stack~\cite{elk} is an application stack for
collecting, indexing, and visualizing information from logs, such as those from
Syslog or HTTP server. Elasticsearch is used for indexing.
Logstash provides network interfaces for receiving logs. Kibana is the web
frontend for searching and visualizing logs. We use \texttt{sebp/elk} for our ELK
stack. Our desired partitions would be one each of Elasticsearch, Logstash, and
Kibana, and one of the startup script.

To exercise this container, we fed it logs from the Linux Audit
subsystem~\cite{rhaudit}. We then performed some simple Elasticsearch queries on
its REST
API interface. Next, we worked with Kibana to create index patterns, ran some
queries on the logs covered by those index patterns, and plotted the query results
on various graphs (this is the visualization part of Kibana). \sysname{} was able
to produce functional partitions as desired. Except for a log file, there is no file
sharing among the partitions; this is expected as the three main components
communicate via network only.

\paragraph{Discussion.}
We note several points from our experience above.
First, container configuration highly influences the degree of isolation
isolation possible among resulting containers. For example, the location of
MySQL socket resulted in the sharing of the entire MySQL data directory with the
HTTPD container in the Mediawiki example. However, this issue was not present
for the Wordpress container, which created the MySQL socket elsewhere. Second,
the different container partitions sometimes have duplicated read-only
resources. For example, the linker and \texttt{libc} are required by all
containers and are currently duplicated across partitions. \sysname{}
could save space by sharing a layer of files common among partitions (Docker
container images consist of read-only layers that can be shared between
containers~\cite{dockerimages}). Finally, as much as 35-67\% of the analysis
time is spent in recovering a file tree from a Docker image and not in actual
analysis. This time can
be saved if the file tree were readily available, such as when using
the Btrfs storage driver~\cite{btrfsdriver}.

\begin{figure}
  \centering
  \includegraphics[width=\vcolumnwidth]{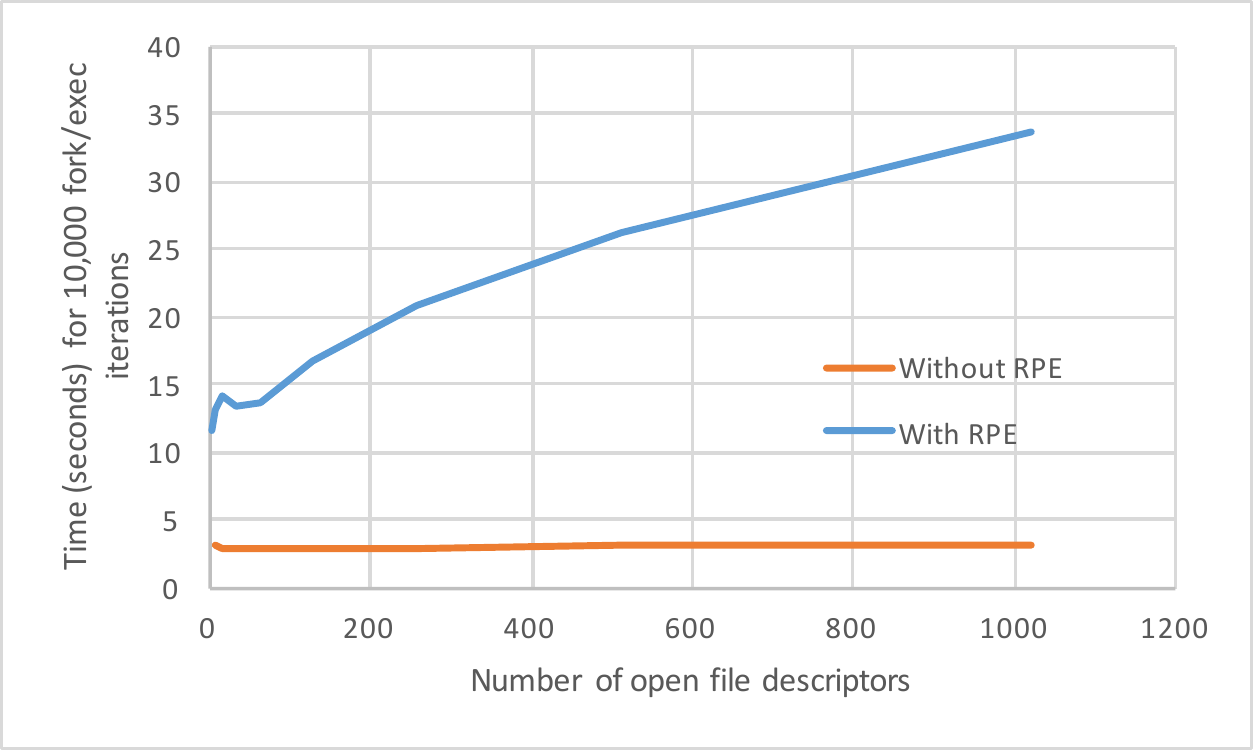}
  \includegraphics[width=\vcolumnwidth]{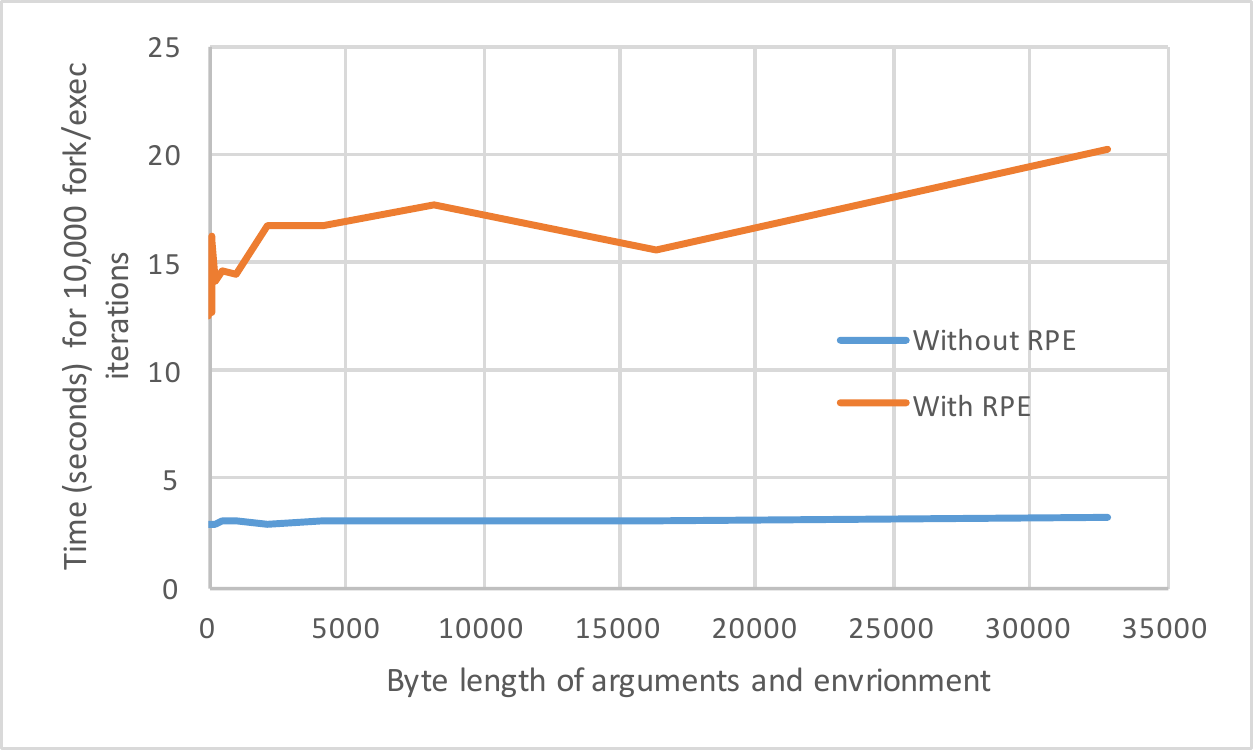}
  \caption{Microbenchmark results showing runtime overhead of RPE as number of
  open file descriptors and arguments length varies}
  \label{fig:microbench}
\end{figure}
\subsection{Runtime Overhead}
We did not find any observable difference in the performance of original and
\sysname-created containers. This is understandable because programs in the
\sysname-created containers execute the same way as the original
containers.
The only overhead incurred is due to the glue of remote process execution but is
likely amortized and becomes negligible.

In order to measure the overhead from RPE
we wrote a microbenchmark consisting of several iterations of
\texttt{fork}/\texttt{exec} calls. \sysname{} results in partitioning at these
calls and adding the glue in between, so we can compare the overhead of the
glue over the original \texttt{fork}/\texttt{exec} setup. Among the artifacts
replicated across containers, the number of open file descriptors and the number
of bytes in the command arguments and environment are variable. We show the
results of our microbenchmark in Figure~\ref{fig:microbench} (obtained on the
same compute platform as that used for our case studies) with both the original and
partitioned setup and with varying open file descriptors and arguments and
environment length. The RPE overhead increases linearly with these variables. This
is intuitive because we replicate these variables in the remote process in
userspace. Nevertheless, programs do not use extremely long command arguments
(for portability reasons), and even if a program opens a large number of files,
the files will typically be closed at the time of \texttt{exec} to prevent file
descriptor leakage issues.  The overhead per \texttt{fork}/\texttt{exec} is thus
only about 1-4 ms for all practical purposes and is easily amortized for
any program that runs for more than a few milliseconds. 

%% file: related.tex
\section{Related Work}\label{sec:related}

A few works in the past have used resource identification for various purposes.
In the Docker ecosystem itself, some blog posts and projects have
developed automatic container slimming as a solution to the big size of Docker
images. All these works~\cite{k15,b15,q15} have relied on
\texttt{fanotify} to identify necessary file resources. This technique is
simpler than system call-based identification but does not record crucial file
system events like creation and moving of files. Indeed, we have observed behavior such as
file moves in real-world containers (such as MediaWiki container that we
examined), which would thus break \texttt{fanotify} solutions. Our system call-based approach is
more complete. Furthermore, we go beyond slimming to
provide container partitioning.

CDE~\cite{ge11} is a tool developed before
application containers; it uses \texttt{ptrace} to identify
file resources needed for running an application and packs
them so as to provide a portable runtime for running the application on any
Linux host.
While our resource identification is similar, we offer a more formal
treatment in a new domain; the other challenges
that we address are unique to our work.

Our work draws its motivation from previous work on least privilege and
privilege separation. Krohn et al.~\cite{kef+05} observe that mainstream operating
systems do not provide easy support for implementing least privilege in programs
despite wide acceptance of the principle. The evolving container ecosystem
also faces this problem, which our work helps address.
Provos et
al.~\cite{pfh03} performed privilege separation of the SSH service manually.
\sysname{} is automatic and so can scale better.
Brumley and Song~\cite{bs04} developed automatic
techniques based on static analysis for automatic privilege separation. Bittau
et al.~\cite{bmhk08} provide operating system-level primitives to allow developing
applications in a privilege-separated manner. Others have used specialized
programming languages and compilers to ensure flow control across program
parts~\cite{ml97,m99,zznm01}. Program partitioning
has also been used in non-security contexts to improve performance~\cite{cmam12}.
Because of the need to analyze real-world containers running on stock Linux, we
have the privilege of neither specialized programming languages nor special
OS-level primitives, nor is our problem very amenable to static analysis
(Section~\ref{sec:approach}).
Nonetheless, these works were valuable in
inspiring our approach to container partitioning and may in the future be used
with our container partitioning approach to offer finer partitions than what we
currently achieve.

Another string of related
works~\cite{vsj13,vgpj14} proposes defenses against resource access attacks that
may happen due to confusion between the intended resource and the actual
resource access. Our work partially mitigates these kinds of attacks as we place
the resources together with the programs in sandboxes protected from outside
attackers.

Remote process execution may be compared to live migration. Live
migration of processes~\cite{do91,ossn02} or virtual machines~\cite{cfh+05}
includes saving all the relevant state and replicating it somewhere else. This
includes transfer of the process or VM's memory as well and usually needs kernel
or hypervisor support. In contrast, RPE is a
light-weight technique to transfer execution right when it begins.
It does not require kernel support but rather takes advantage of a shared
kernel state to enable low-overhead, transparent remote execution.

%% file: discussion.tex
\section{Limitations}\label{sec:discussion}

\sysname{} provides an important first step in container partitioning and
slimming.
In this section, we point out limitations of our approach and discuss
directions for future research. \sysname{} inherits the usual drawbacks of
dynamic
analysis: if test runs do not cover all relevant scenarios, resource
identification will be incomplete.
There are at least two promising future directions to mitigate this limitation:
($a$) design an easy-to-use interface through which a user can guide
our tool as to which resources will be necessary; and ($b$) conservatively
expand the resource set to be placed in a container based on patterns seen in
resources already being used. Considering a simple example for ($b$), if a
container meant for compiling Latex documents uses *\texttt{.sty} under
\texttt{/usr/local/texmf}, a regular expression learning algorithm~\cite{a87}
could be used to deduce this pattern to automatically add all
*\texttt{.sty} under this tree, even if they are not used in test runs.

A second limitation is that system calls cannot be used for identifying and
partitioning some kinds of resources. Environment variables are one example of
resources that may be considered worth partitioning but are not managed through
system calls. For such resources, static analysis or an information flow-tracing
dynamic analysis may be used to augment our approach.

%% file: conclusion.tex
\section{Conclusion}\label{sec:conclusion}

Application containers, such as those provided by Docker, are becoming
increasingly popular.
We however observe that many Docker containers violate the principle of least
privilege by including more resources than are necessary and by not providing
privilege separation between applications running in the same container. We
designed and implemented \sysname{}, which partitions a container into many simple ones,
enforcing privilege separation between them, and eliminates the resources that
are not necessary for application execution. To achieve these ends, we developed
techniques for identifying resource usage, for performing partitioning, and for
gluing the partitions together to retain original functionality. Our evaluation shows that
\sysname{} creates functionality-preserving partitions, achieves
container image size reductions of 58-95\% and processes even big containers in
less than thirty seconds.